\documentclass[prd,superscriptaddress]{revtex4}

\usepackage{epsfig}
\usepackage{empheq}
\usepackage{xcolor}

\newcommand{\PRE}[1]{}
\newcommand{\be}{\begin{equation}}
\newcommand{\ee}{\end{equation}}
\newcommand{\bea}{\begin{eqnarray}}
\newcommand{\eea}{\end{eqnarray}}

\newcommand{\ba}{\begin{array}}
\newcommand{\ea}{\end{array}}

\newcommand{\mat}[2][ccccc]{\left( \begin{array}{#1} #2\\ \end{array}\right)}

\newcommand{\lsim}{\mathrel{\hbox{\rlap{\hbox{\lower4pt\hbox{$\sim$}}}\hbox{$<$}}}}
\newcommand{\gsim}{\mathrel{\hbox{\rlap{\hbox{\lower4pt\hbox{$\sim$}}}\hbox{$>$}}}}
\newcommand{\til}{\widetilde}

\newcommand{\tev}{\text{TeV}}
\newcommand{\gev}{\text{GeV}}

\begin{document}

\title{
\PRE{\vspace*{1.5in}}
Neutrino dark matter candidate in fourth generation scenarios
}
\author{Hye-Sung Lee}
\affiliation{Department of Physics, Brookhaven National Laboratory, Upton, NY 11973, USA}
\author{Zuowei Liu}
\affiliation{C.N. Yang Institute for Theoretical Physics, Stony Brook University, Stony Brook, NY 11794, USA}
\author{Amarjit Soni}
\affiliation{Department of Physics, Brookhaven National Laboratory, Upton, NY 11973, USA}
\date{May 2011}

\begin{abstract}
\PRE{\vspace*{.1in}} \noindent
We overview the constraints on the 4th-generation neutrino dark matter candidate and investigate a possible way to make it a viable dark matter candidate.
Given the LEP constraints tell us that the 4th-generation neutrino has to be rather heavy ($>M_Z/2$), in sharp contrast to the other
three neutrinos, the underlying nature of the 4th-generation neutrino is expected to be different.
We suggest that an additional gauge symmetry $B-4L_4$ distinguishes it from the Standard Model's three lighter neutrinos and this also facilitates promotion of the 4th-generation predominantly right-handed neutrino to a good cold dark matter candidate.
It provides distinguishable predictions for the dark matter direct detection and the Large Hadron Collider experiments.
\end{abstract}

\maketitle

\section{Introduction}
Dark matter (DM) is one of the best evidences for physics beyond-the-standard-model (BSM).
DM consists of the 22\% of the energy budget of the Universe, which is much more than that of the entire standard model (SM) particles.
Thus, any BSM better have a good DM candidate.

The 4th-generation (4G) scenario is one of the well-motivated BSMs.
We have already seen three generations; the 4G may exist as well.
Besides, the heavy 4G fermion masses may lead to dynamical electroweak (EW) symmetry breaking \cite{Holdom:1986rn,Carpenter:1989ij,King:1990he,Hill:1990ge,Burdman:2007sx,Hung:2009hy,Hashimoto:2009ty}.
Furthermore, whereas $CP$ violation from the SM, i.e. the phase in the CKM matrix, is not large enough for successful EW baryogenesis, addition of another generation opens up an important new avenue in this context \cite{Hou:2008xd}.
There is also an argument that an even number of fermion generations is more natural from the string theory point of view \cite{Cvetic:2001nr,Lebed:2011zg}.
Lastly, there are also several anomalies, in particular, in $CP$ violation in $B$-physics \cite{Lunghi:2007ak,Lunghi:2008aa,Lenz:2010gu,Bona:2009cj,Lunghi:2010gv}, which can be addressed simply by adding another generation of quarks \cite{Soni:2008bc,Nandi:2010zx,Buras:2010pi,Hou:2010mm}.

Indeed one of the appealing reasons to consider the 4G is because it is one of the simplest ways to extend the SM.
We can compare this to many other new physics scenarios including supersymmetry, extra dimension, and little Higgs whose structure is much more complicated than simple addition of a fermion generation.
In this context, it is important to note that whereas in practically all BSMs the absence of flavor-changing neutral current (FCNC) (say in $b \to s \gamma$) contributions beyond that of the SM is always a problem, the 4G easily accounts for this~\cite{Lunghi:2011xy}. 

For these reasons the 4G scenario is recently getting increasing attention.
(For some examples, see Refs.~\cite{DePree:2010zz,Alwall:2010jc,Erler:2010sk,He:2010nt,Kong:2010qd,Carpenter:2010dt,Eberhardt:2010bm,Das:2010fh,Chanowitz:2010bm,Dawson:2010jx,Ishiwata:2011ny,Anastasiou:2011qw,Soni:2010xh,Keung:2011zc,Lenz:2011gd,Atwood:2011kc,Masina:2011ew,Cotta:2011bu,Smith:2011rp,Barger:2011tb}.)
Thus, it is of great interest to develop natural DM candidates in the 4G models.
The 4G neutrino actually has some characteristics to be a good DM candidate, e.g, electrically neutral and massive, but it does not satisfy all experimental constraints to be a viable DM candidate.
In this paper, we will review the constraints on the 4G neutrino DM candidate and investigate a possible way out of these constraints.

The outline of the rest of this paper is given as following.
In Section \ref{sec:constraints}, we describe the constraints on the neutrino DM candidate.
In Section \ref{sec:approach}, we describe our idea and approach.
In Section \ref{sec:phenomenology}, we discuss some of the phenomenology of the model.
In Section \ref{sec:summary}, we summarize our result.
In Appendix \ref{sec:appendix_relic}, we describe a detail calculation of the neutrino DM relic density.

\section{Constraints on the neutrino dark matter candidate}
\label{sec:constraints}
\begin{figure}[t]
\begin{center}
\includegraphics[height=3cm]{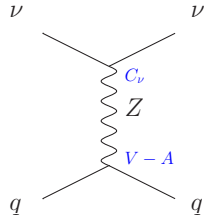} 
\end{center}
\caption{Direct detection of a massive neutrino dark matter candidate through the SM $Z$ boson. For the Dirac neutrino ($C_\nu = V-A$) case, the spin-independent cross section (via $V$ and $V$ interaction) dominates.
For the left-handed Majorana neutrino ($C_\nu = A$) case, the spin-dependent cross section (via $A$ and $A$ interaction) dominates.}
\label{fig:direct1}
\end{figure}

One of the most stringent constraints on the 4G neutrinos comes from the invisible $Z$ decay width.
The partial decay width of the $Z$ boson to light active neutrinos is given by $\Gamma (Z \to \bar\nu \nu) = N_f \times 0.17 ~\gev$ with $N_f$ being the number of generations, and the LEP experiments tells us there are only three light active neutrino flavors \cite{Nakamura:2010zzi}.
Thus, the 4G active neutrinos (i.e. those with significant left-handed neutrino component) are constrained by, $m_\nu \gsim M_Z/2$, and this is an important issue in all 4G models.
(See, for example, Ref.~\cite{King:1992qr}.)
Since we consider a 4G neutrino as a DM candidate, there are more constraints.
A viable thermal DM candidate should (i) be cold, neutral, stable, (ii) satisfy the measured DM relic density constraint \cite{Komatsu:2010fb}, and (iii) satisfy the direct DM detection bounds \cite{Angle:2008we,Ahmed:2009zw,Aprile:2011hi}.

To satisfy the WMAP data, the DM relic density should be $\Omega h^2 \sim 0.1$ or smaller if there are multiple DMs.
For a cold relic neutrino ($m_\nu \gg 1~\text{MeV}$), the relic density requires $m_\nu \gsim {\cal O} (\gev)$ for a Dirac neutrino, and the bound is not too different (only several times larger) for a Majorana neutrino \cite{Lee:1977ua,Kolb:1985nn}.

The direct detection constraint on the heavy neutrino DM depends on the type of neutrino.
For the 4G Dirac neutrino DM, there is a spin-independent (or coherent) interaction from the $V$-$V$ coupling through the SM $Z$ boson (see Fig.~\ref{fig:direct1}).
The cross section per nucleon is given as
\be
\sigma_{\nu - \text{nucleon}}^{SI} = \frac{1}{\pi} m_\text{eff}^2 \left( \frac{Z f_p + (A-Z) f_n}{A} \right)^2 \sim 0.1 G_F^2 m_\text{eff}^2 \sim 10^{-38} ~\text{cm}^2
\ee
where $m_\text{eff}=\frac{m_{\nu}m_{p}}{m_{\nu}+m_{p}}$ with $m_{\nu}$ being the neutrino DM mass and $m_{p}$ being the proton mass, and the couplings $f_p = \frac{G_F}{2\sqrt{2}} (1-4\sin^2 \theta_W)$ for proton and $f_n = - \frac{G_F}{2\sqrt{2}}$ for neutron.
To satisfy the current CDMS and XENON spin-independent cross section bounds (which is roughly $10^{-44} ~\text{cm}^2$ level for the EW scale DM) \cite{Ahmed:2009zw,Aprile:2011hi}, the neutrino mass should be many orders of magnitude larger than the EW scale. 

For the 4G Majorana neutrino DM, there is only a spin-dependent interaction from the $A$-$A$ coupling since the vectorial coupling of a Majorana particle to a vector boson is always zero (see Fig.~\ref{fig:direct1}).
In supersymmetry (SUSY) models, the neutralino (a Majorana fermion) DM candidate does not couple to $Z$ through a vectorial coupling for the same reason.
The cross section is given as
\be
\sigma_{\nu - \text{nucleon}}^{SD} = \frac{8 G_F^2}{\pi} m_\text{eff}^2 \left( a_p \left< S_p \right> + a_n \left< S_n \right> \right)^2 \frac{J+1}{J}
\ee
where the effective DM-nucleon couplings $a_{p,n}$ and the expectation values of the spin content of the nucleon within the nucleus $\left< S_{p,n} \right>$ are inputs from nuclear physics.
$J$ is a total nuclear spin.
The XENON10 collaboration provided the interpretation of their results in terms of the Majorana neutrino mass bound: $m_\nu \gsim 2 ~\tev$ \cite{Angle:2008we} for the standard model couplings. 
(See their paper for the detailed assumptions).

Summing up all of direct experimental constraints, we can see that a 4G neutrino DM candidate (no matter it is Dirac or Majorana) should be very heavy to be consistent with  the direct detection constraint because of the expected large cross section through a $Z$ boson.
(For an instance of earlier models with less constraints, see Ref.~\cite{Volovik:2003kh}.
There are also other indirect constraints on the neutrino DM candidates, which is beyond the scope of our paper.
For an instance among early works, see Ref.~\cite{Fargion:1994me}.)
Assuming the 4G neutrino is the sole DM candidate, this would be hard to accommodate without violating {\it perturbative} unitarity.
This is a major difficulty in considering the 4G neutrino as a DM candidate, in addition to explaining why the 4G neutrino is heavy and stable.
In the following section, we investigate what we need in order to make a 4G neutrino a valid DM candidate and propose a possible approach that simultaneously attempts to address these twin difficulties.

\section{Our approach}
\label{sec:approach}
We start by observing a similar history in the scalar neutrino (sneutrino) DM candidate in the SUSY case.
In earlier days, the sneutrino (at that time the left-handed only) was one of the lightest supersymmetric particle (LSP) DM candidates along with the popular neutralino LSP.
In 1994, it was excluded as a viable DM candidate after it was found that the sneutrino mass range that is consistent with the DM relic density bound cannot satisfy the direct detection constraint \cite{Falk:1994es}.
The spin-independent direct detection cross section would be $\sigma_{\til \nu - \text{nucleon}}^{SI} \sim G_F^2 m_\text{eff}^2 \sim 10^{-37} ~\text{cm}^2$, which was larger than the experimental bound even at that time.

However, it was learned that if a rather (predominantly) right-handed sneutrino is the LSP, and if there is a new massive gauge boson $Z'$ that couples to the right-handed sneutrino, it can be a good LSP DM candidate in SUSY models \cite{Lee:2007mt}.
(For other possible ways to realize right-handed sneutrino as a thermal DM candidate, see, for instance, Refs.~\cite{Cerdeno:2008ep,Cerdeno:2009dv}.)
This became possible because the major channel for the right-handed sneutrino is through the $Z'$ gauge boson whose mass and coupling are different from those of the SM $Z$ boson.
Since we have a similar problem with the left-handed neutrino in the 4G model, which is basically a severe constraint coming from the fact that the SM $Z$ boson is a major channel, we employ a similar approach.
We will introduce a new gauge symmetry and take the 4G (predominantly) right-handed neutrino as the DM candidate.

It might not seem appealing to introduce more than one new physics.
Nevertheless, the reality is that most of new ideas need an auxiliary symmetry to be phenomenologically viable with issues such as EW precision, proton stability, etc.
SUSY needs $R$-parity or an additional $U(1)$ gauge symmetry such as $U(1)_{B-L}$ or $U(1)_{B-x_i L}$ \cite{Lee:2010hf}.
Extra dimension needs $KK$-parity \cite{Georgi:2000ks} and/or some custodial symmetry.
Little Higgs also needs $T$-parity \cite{Cheng:2003ju} or something similar. 
So it should be considered reasonable to introduce an additional $U(1)$ gauge symmetry as an auxiliary symmetry for the 4G models.
Typically, the auxiliary symmetries provide stability to the DM candidates as well (LSP by $R$-parity, LKP by $KK$-parity, LTP by $T$-parity).
In fact, a new symmetry opens a possible avenue to address simultaneously the issue of the heaviness of a 4G neutrino compared to the three neutrinos of the SM.

Now, we will discuss more about the most important features of our approach: (i) an auxiliary gauge symmetry $U(1)_{B-4L_4}$, and (ii) a right-handed Majorana neutrino DM candidate.

\begin{figure}[t]
\begin{center}
\includegraphics[height=3cm]{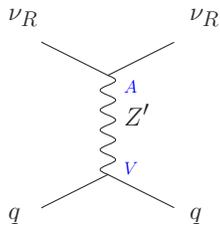} 
\end{center}
\caption{Direct detection of a massive right-handed Majorana neutrino dark matter candidate through $Z'$ boson of the $U(1)_{B-4L_4}$. It predicts null results for both the spin-independent and spin-dependent cross sections.}
\label{fig:direct2}
\end{figure}

\subsection{New gauge symmetry}
Here, we investigate the possibility of an extra Abelian gauge symmetry which has generation-dependent charges in the lepton sector.
We consider a $B-x_i L$ type of model, where $B$ ($L$) stands for the baryon (lepton) number and $x_i$ is a generation-dependent constant.
This type of model was studied for the usual three generation models in Refs.~\cite{Ma:1997nq,Lee:2010hf}.
While the usual lore says $B-L$ is the only possible anomaly-free Abelian gauge extension unless one adds extra fermion fields, it should be noted that this statement is true only for the generation-independent charges case.
With our generation-dependent $U(1)$ gauge symmetry, we may not need exotic fermions that are charged under the SM gauge interactions.

Because of the vectorial nature of $SU(3)_C$ and $U(1)'$, the $[SU(3)_C]^2$-$U(1)'$, $[U(1)']^3$, and $[\text{gravity}]^2$-$U(1)'$ anomalies are automatically zero.
The sum of hypercharges is zero separately for each family of quarks and for each family of leptons, and it makes the $[U(1)']^2$-$U(1)_Y$ anomaly vanish.
Thus we need to consider only the remaining two anomaly conditions similarly to the aforementioned \cite{Ma:1997nq,Lee:2010hf}.
\begin{eqnarray}
\mbox{$[SU(2)_L]^2$-$U(1)'$} &:& N_f ~(3) (1/3) + \sum_{i=1}^{N_f} (-x_i) = 0  \\
\mbox{$[U(1)_Y]^2$-$U(1)'$} &:& N_f ~(3)[2(1/6)^2-(2/3)^2-(-1/3)^2](1/3) + [2(-1/2)^2-(-1)^2] 
\sum_{i=1}^{N_f} (-x_i) = 0 
\end{eqnarray}
Both are satisfied if
\begin{equation}
\sum_{i=1}^{N_f} x_i = N_f
\end{equation}
where $N_f = 4$ is the number of total fermion generations.
We want $x_k (\text{for}~ k=1,2,3) = 0 \ne x_4$
to keep the regular seesaw mechanism only for the SM neutrinos and avoid it for the 4G neutrinos.
This leads to
\be
x_1 = x_2 = x_3 = 0, \quad x_4 =4
\ee
which means that the charge of our new $U(1)$ model should be
\be
{\cal Q} = B - 4 L_4 .
\ee
This is a similar form to the aforementioned $B-3L_\tau$ of a typical three generation model in Ref.~\cite{Ma:1997nq} except that it is the 4G leptons that have nonzero charges.
The charge is $B=1/3$ for all the quarks, $0$ for the SM leptons, and $-4L_4 = -4$ for the 4G leptons.
Thus, the charge is effectively the baryon number ($B$) for the SM particles.

\subsection{Right-handed neutrino dark matter candidate}
In our model, the 4G right-handed neutrino ($N_4$) has a non-zero $U(1)'$ charge ${\cal Q}[N_4] = -4$, and the Majorana neutrino mass term would require $S N_4 N_4$ type of term where $S$ is a Higgs singlet that can break the $U(1)'$ gauge symmetry spontaneously.
It fixes ${\cal Q}[S] = -2 {\cal Q} [N_4] = 8$.
Then 4G neutrino lagrangian would be
\begin{equation}
{\cal L} \sim y_D L_4 H N_4 + y_S S N_4 N_4
\label{eq:simplest}
\end{equation}
which gives the 4G neutrino mass matrix of $m_\nu \sim \mat{0 & y_D v_D \\ y_D v_D & y_S v_S}$ with $v_D = \left< H \right>$ and $v_S = \left< S \right>$.
The seesaw mechanism would work for $y_D v_D < y_S v_S$, which makes the lighter neutrino always possess significant left-handed component that couples to the SM $Z$ boson.
Thus, in order to have the (predominantly) right-handed neutrino as the lightest 4G neutrino, we need to change the mass matrix.

For one possibility, one can introduce a Higgs triplet $T$ (or more) that can give a mass term to the left-handed neutrinos.
It would allow
\begin{equation}
{\cal L} \sim y_T T L_4 L_4 + y_D L_4 H N_4 + y_S S N_4 N_4
\label{eq:snn}
\end{equation}
which gives the mass matrix of $m_\nu \sim \mat{y_T v_T & y_D v_D \\ y_D v_D & y_S v_S}$ with $v_T = \left< T \right>$.
(See Ref.~\cite{Ma:1998dr} to see some details in using Higgs triplets in $B-3L_\tau$ model.)
In this case, we would need a condition $y_T v_T > y_S v_S$ and $y_D v_D \simeq 0$ to ensure that the lightest 4G neutrino is predominantly right-handed. 
(Though a detailed analysis would be called for, such an EW scale vacuum expectation value of the Higgs triplet could be still consistent with the EW precision data, due to the possible cancellations between the Higgs and the 4G fermions in computing the oblique corrections \cite{Kribs:2007nz,He:2001tp}.)
In our analysis, we will work in the limit in which the mixing between the 4G left-handed neutrino and the 4G right-handed neutrino is negligible when forming the mass eigenstates. 

It is conceivable to introduce a $Z_2$ parity which gives odd parity only for 4G leptons in order to ensure the lightest 4G lepton is stable.
Since the lightest neutrino would be predominantly right-handed, it might be stable enough to be a good DM candidate without the parity though.
The $N_4$ would not decay to the SM particles through the Yukawa term $L_\text{SM} H N_4$ since the Yukawa term is forbidden by the $B-4 L_4$ symmetry.
The nonrenormalizable effective Yukawa terms $(S/\Lambda)^n L_k H N_4$ (for $k=1,2,3$) and $(\bar H H /\Lambda^2)^n L_k H N_4$ are also forbidden since their total charges are $8n-4$ and $-4$, respectively.

There are other possibilities for the neutrino masses including the extra dimensions \cite{ArkaniHamed:1998vp}.
Since the details of realization of the right-handed neutrino as the lightest 4G neutrino and its associated phenomenology are beyond the interest of our current study, we do not discuss them further in this paper.

\section{Discussion on the 4th-generation neutrino dark matter}
\label{sec:phenomenology}
\begin{figure}[tb]
\begin{center}
\includegraphics[height=5cm]{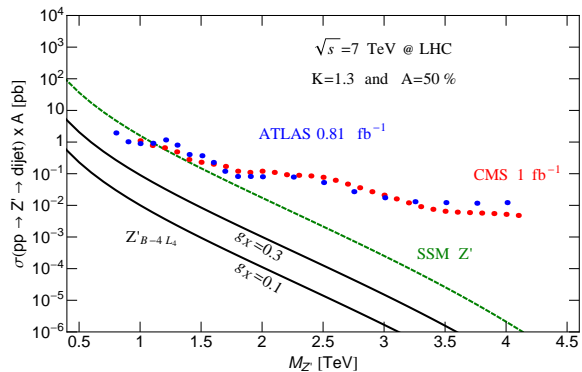} 
\end{center}
\caption{The constraints on $Z'$ in the dijet search from the CMS \cite{CMS} data (red dots) with integrated luminosity of 1 fb$^{-1}$ and from the ATLAS \cite{ATLAS} data (blue dots) with integrated luminosity of 0.81 fb$^{-1}$.
The predicted events at LHC through the $Z'$ boson of the $U(1)_{B-4L_4}$ with a range of coupling constant (for instance, $g_X=0.3$) lie below the current reach of the CMS and ATLAS data.
Also plotted are the predicted signals from the sequential $Z'$ model (green dashes) for the comparison.}
\label{fig:collider}
\end{figure}

The numerical calculation of the 4G right-handed neutrino relic density is given in Appendix \ref{sec:appendix_relic}.
As it is detailed there, though the 4G right-handed neutrino couples only to the $Z'$ gauge boson, the right relic density can be easily achieved using the $Z'$ resonance.
In this sense, our model predicts that the mass of the 4G right-handed neutrino is close to $M_{Z'}/2$. 
Since the 4G fermion masses cannot be much larger than EW scale because of perturbative unitarity and the 4G right-handed neutrino should be the lightest among the 4G leptons to be a good DM candidate, it implies the $Z'$ should be EW/TeV scale, which is interesting for the Large Hadron Collider (LHC) phenomenology. 
The direct detection cross section for the 4G right-handed neutrino DM candidate is predicted to be negligible.
The Majorana neutrino has only an axial-vector coupling to $Z'$ and the quarks have only vector couplings to $Z'$ in our model.
Thus, our model predicts the null results for the spin-independent and spin-dependent direct detection cross sections in the nonrelativistic limit (see Fig.~\ref{fig:direct2}). 

As we look back, we can see all the constraints are satisfied in our model.
(1) Invisible $Z$ width is satisfied as the 4G left-handed neutrino is heavy.
(2) The 4G right-handed neutrino DM candidate is massive, neutral and stable.
(3) The right relic density can be achieved through the $Z'$ resonance.
(4) The direct detection bound is satisfied as our model predicts null result for both spin-independent and spin-dependent cross sections. 
Thus, the 4G right-handed neutrino in our model is a viable DM candidate and an additional $U(1)$ gauge symmetry ($B-4L_4$) is well motivated as an auxiliary symmetry for the 4G scenarios.

We have not discussed aspects on the collider experiments.
Though it is not our purpose of this paper to study them in detail, there are some immediate observations one can make for the LHC experiments in our model.
(1) Null result for the typical dilepton resonance ($\bar q q \to Z' \to \bar\ell \ell$) since ${\cal Q}[\ell_\text{SM}] = 0$ (practically leptophobic $Z'$).
(2) Dijet resonance ($\bar q q \to Z' \to \bar q q$) would be observable if they can be distinguished from the backgrounds, since $Z'$ couples to quarks.
(3) New 4G particle production channels ($\bar q q \to Z' \to \bar q_4 q_4$, $\bar e_4 e_4$) are available. 

The dijet resonance searches already provide some limits for the couplings \cite{Aaltonen:2008dn,CMS,ATLAS}.
Fig.~\ref{fig:collider} shows our estimated dijet $Z'$ resonance cross section at the LHC experiments with $\sqrt{s} = 7 ~\tev$.
For our estimation, we used narrow width approximation with the higher-order correction $K$-factor $K = 1.3$ and acceptance $A = 0.5$.
The estimation shows that our model with a wide range of coupling constant (for instance, $g_X = 0.3$) is not ruled out by the current LHC data of dijet resonance search at the CMS and the ATLAS (with integrated luminosity of 1 fb$^{-1}$ and 0.81 fb$^{-1}$, respectively).

When an auxiliary symmetry is introduced, it changes the predictions of the new physics.
For instance, major SUSY discovery channels are significantly different depending on the presence of the $R$-parity or some $U(1)$ gauge symmetries that are introduced for the proton stability in the SUSY models \cite{Lee:2008cn,Lee:2010hf}.
In our model, the $B-4L_4$ symmetry allows large production of the 4G fermions through the on-shell resonance of the $Z'$ at the LHC.
Considering that the off-shell gluon channel ($\bar q q \to g^* \to \bar q_4 q_4$) is the major 4G production channel in the usual 4G scenarios, we can see the chance to observe the 4G signal is much larger in the presence of the auxiliary symmetry $U(1)_{B-4L_4}$.
Since ${\cal Q}[H] = 0$, any signals from the $Z' \to Z + H$ at the LHC such as discussed in Refs.~\cite{Barger:2009xg,Katz:2010mr} would be absent.

\section{Summary and Outlook}
\label{sec:summary}
We reviewed the constraints on the 4G neutrino DM candidate, 
and developed a possible way to get around it.
It is shown that a 4G right-handed neutrino that couples to the EW/TeV scale $Z'$ can be a good DM candidate.
This motivates an additional gauge symmetry $U(1)_{B-4L_4}$ as an auxiliary gauge symmetry of the 4G scenarios.
Among the leptons, only the 4G ones are charged under the new gauge symmetry, and it explains relevant constraints from various data including the LEP $Z$ width measurement, DM direct detection experiments, and DM relic density measurement.
Especially, it predicts that the typical DM direct detection experiments would not see any signal, casting a need to develop new methods to detect this kind of DM.

In our picture, the matter energy budget of our Universe might be dominated by the 4G.
The 4G scenario not only predicts new fermions, but also predict a new flavor-dependent gauge symmetry at EW/TeV scale, which leads to distinguishable predictions for the LHC phenomenology.

\acknowledgments
We thank Hooman Davoudiasl, Wai-Yee Keung, and Kai Wang for useful discussions.
This research was supported in part by the DOE grant No. DE-AC02-98CH10886 (BNL) and by the NSF grant PHY-0969739 (Stony Brook).

\appendix

\begin{figure}[tb]
\begin{center}
\includegraphics[height=5cm]{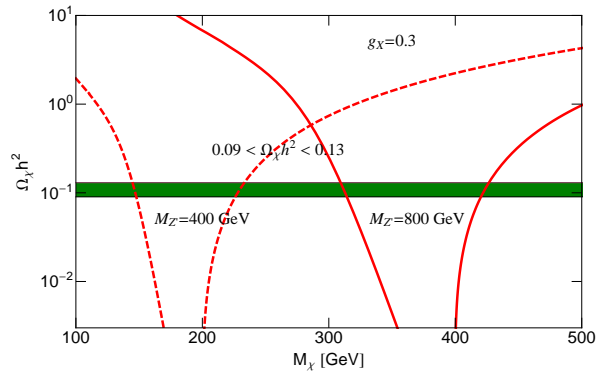} 
\end{center}
\caption{The relic density of the dark matter vs. the dark matter mass for $M_{Z'} = 400$ GeV (dashed) and $800$ GeV (solid) with a coupling constant $g_X=0.3$.
The green band is the WMAP measured dark matter relic density.}
\label{fig:omega}
\end{figure}

\section{Relic density of the 4th-generation neutrino dark matter candidate}
\label{sec:appendix_relic}
Here we discuss the detailed calculation of the relic density of the 4G right-handed neutrino DM candidate.
Under $B-4L_4$ gauge symmetry, the $Z'$ gauge boson does not couple to the first three generation leptons.
The relevant Lagrangian is given by
\begin{equation}
{\cal L}^{Z'}_\text{int}= g_X Z'_{\mu} \left[   \frac{1}{6}  \bar q\gamma^{\mu} q 
-2  \bar e_4 \gamma^{\mu} e_4
 -2 \bar{\nu}_{4L}\gamma^{\mu} \gamma_5 \nu_{4L} -2 \bar{\nu}_{4 R} \gamma^{\mu} \gamma_5 \nu_{4 R} \right]
\label{eq:lagrangian}
\end{equation}
where $g_X$ is the coupling constant, $q$ are the quarks including the 4G ones, $e_4$ is the charged lepton in the 4G, $\nu_{4L}$ and $\nu_{4R}$ are the 4G Majorana neutrinos. 
Above, we have neglected a possible small mixing between the chiral neutrinos as 
discussed before. 
We will adopt the typical DM notation $\chi\equiv \nu_{4 R}$.

The decay widths of the $Z'$ boson into the quarks and the 4G leptons are
\begin{empheq}{align}
\Gamma(Z'\to \bar q q) & = N_C \frac{M_{Z'}}{12\pi}\bigg[\frac{g_X}{6}\bigg]^2
\sqrt{1-4\frac{M_q^2}{M_{Z'}^2}}
\bigg(1+2\frac{M_q^2}{M_{Z'}^2}\bigg)
\\ \Gamma(Z'\to \bar e_4 e_4) & = \frac{M_{Z'}}{12\pi}\bigg[2 g_X\bigg]^2
\sqrt{1-4\frac{M_{e_4}^2}{M_{Z'}^2}}
\bigg(1+2\frac{M_{e_4}^2}{M_{Z'}^2}\bigg)
\\  \Gamma(Z'\to \bar \nu_{4L} \nu_{4L})  & =  \frac{1}{2}\frac{M_{Z'}}{12\pi} \bigg[2 g_X\bigg]^2
\bigg[ 1-4\frac{M_{\nu_{4L}}^2}{M_{Z'}^2}\bigg]^{3/2} 
\\ \Gamma(Z'\to \bar \chi \chi)  & =  \frac{1}{2}\frac{M_{Z'}}{12\pi} \bigg[2 g_X\bigg]^2
\bigg[ 1-4\frac{M_{\chi}^2}{M_{Z'}^2}\bigg]^{3/2} 
\end{empheq}
where $N_C=3$.

\begin{figure}[tb]
\begin{center}
\includegraphics[height=5cm]{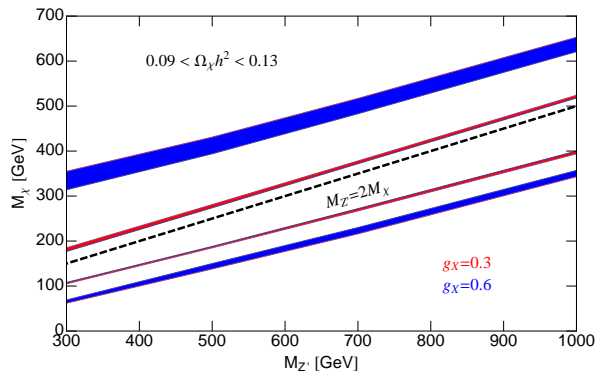} 
\end{center}
\caption{The black-dashed line corresponds to the relation $M_{Z'}=2M_{\chi}$. 
The red (blue) bands are where the condition $0.09<\Omega_{\chi}h^2<0.13$ is satisfied when $g_X=0.3~(0.6)$.
We assume that the other particles in the fourth generation are heavier so that they do not contribute to the $Z'$ decay widths and dark matter thermal annihilation cross section.}
\label{fig:mass}
\end{figure}

The annihilation cross section for the process $\chi\chi \to Z' \to \bar q q$ is given as 
\begin{equation}
\sigma v_{\rm rel}(\chi\chi\to\bar qq)  =  
 \frac{g^4_X}{12\pi } \frac{s\beta^2_{\chi} \beta_q \big[ 1-\beta^2_q/3 \big]}
{(s-M^2_{Z'})^2+M^2_{Z'}\Gamma^2_{Z'}}  
\end{equation}  
where 
$v$ is the relative velocity between colliding DM particles, 
$s=16M^2_{\chi}/(4-v^2)$, $\beta_q = \sqrt{1-4M^2_q/s}$, and 
$\beta_{\chi} = \sqrt{1-4M^2_{\chi}/s} = v/2$. 

For definiteness, we assume the final states are the SM quarks (all three generations) only, although it is in principle possible to have the 4G quarks lighter than the 4G right-handed neutrino DM candidate.
The SM leptons do not carry the $U(1)_{B-4L_4}$ charge and cannot be the final states in the $Z'$ mediated annihilation.

The DM relic density at present is given by (see e.g. Ref.~\cite{Griest:1990kh})
\begin{equation}
\Omega_\chi h^2 \approx \frac{1.07 \times 10^9~{\rm GeV^{-1}}}{J \sqrt{g_{*}}M_{\rm Pl}}
\end{equation}
where $g_{*}$ is the number of degrees of freedom of the relativistic particles at the time of 
freeze-out, $M_{\rm Pl}=1.22 \times 10^{19}$ GeV is the 
Planck mass.
The quantity $J$ is given as  
\begin{equation}
J = \int_{0}^{x_f} {\rm d}x \langle \sigma v\rangle
= \int_{0}^{x_f} {\rm d}x \int_{0}^{\infty} {\rm d}v \, 
\frac{ v^2 e^{-v^2/4x}}{2\sqrt{\pi x^3}} \sigma v(\chi\chi \to {\rm all})
~~ {\rm with} ~~
x \equiv \frac{T}{M_{\chi}}
\end{equation}
and $x_f=T_f/M_{\chi}$ is the reduced freeze-out temperature. 

In order to generate the right amount of DM, the annihilation should occur near the vicinity of the $Z'$ pole.
In Fig.~\ref{fig:omega}, we present for two $Z'$ masses $M_{Z'}=400$, $800$ GeV with a coupling constant $g_X=0.3$.
It shows that in order to satisfy the relic density constraint from WMAP, $0.09<\Omega_\chi h^2 <0.13$ \cite{Komatsu:2010fb}, which is indicated by the green band, the DM mass has to be close to a half of $Z'$ mass.
The relic density can be satisfied on both sides of the $Z'$ resonances.
For some recent works on DM annihilation via $Z'$ boson, see e.g. Refs.~\cite{Feldman:2007wj,Lee:2007mt,LUP}. 

We further illustrate the relic density constraint on the mass relations between the DM and the $Z'$ boson in Fig.~\ref{fig:mass} for the coupling constant $g_X = 0.3$ and $0.6$.
The red (blue) bands for $g_X = 0.3$ ($0.6$) are plotted for the relic density in the range $0.09<\Omega_{\chi}h^2<0.13$.
The upper edges of the blue and red bands that are above the $M_{Z'} = 2 M_\chi$ line (black dashed) correspond to $\Omega_{\chi}h^2=0.13$. 
For the two bands below the $M_{Z'} = 2 M_\chi$ line, the upper edges of the bands correspond to $\Omega_{\chi}h^2=0.09$. 
Fig.~\ref{fig:mass} shows that the decrease in the magnitude of the gauge coupling $g_X$ from $g_X=0.6$ to $g_X=0.3$ 
results in the shift of the DM mass $M_{\chi}$ towards the resonance.
The DM has to annihilate closer to the resonance for a smaller coupling constant in order to maintain the cross section that produces the right amount of DM. 
Fig.~\ref{fig:mass} also shows that the DM annihilates closer to the resonance on the right hand side of the pole, which are indicated by the bands below the $M_{Z'} = 2 M_\chi$ line, than on the left hand side of the pole which are indicated by the bands above the $M_{Z'} = 2 M_\chi$ line.

\begin{figure}[tb]
\begin{center}
\includegraphics[height=5cm]{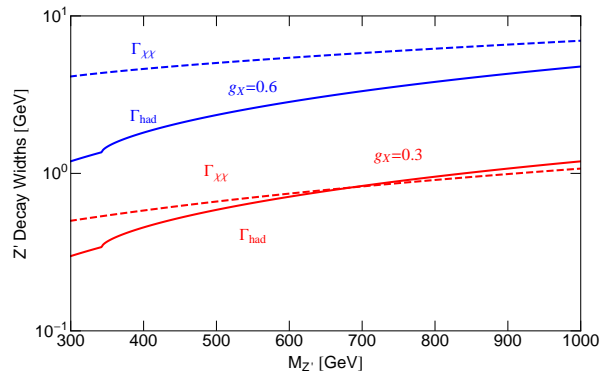} 
\end{center}
\caption{Decay widths of the $Z'$ boson for $g_X=0.3$ (red) and $g_X=0.6$ (blue).
$\Gamma_{\chi\chi}$ is the decay width to the 4G right-handed neutrino dark matter final state, and $\Gamma_\text{had}$ is the sum of all SM quarks contribution.
In the plot, the mass of the right-handed neutrino is chosen to satisfy the relic density constraint in Fig.~\ref{fig:mass} and the $Z'$ can decay into its pair.}
\label{fig:zprime}
\end{figure}

In Fig. \ref{fig:zprime}, we exhibit the relevant decay widths of the $Z'$ boson.
The mass relation between the DM and the $Z'$ boson is constrained by the relic density of the DM, which is indicated by the mass bands in Fig. \ref{fig:mass}.
Since the $Z'$ has a relatively larger coupling to the 4G leptons than the quarks in our model, the DM branching ratio can be quite large and comparable to the sum of the hadronic branching ratios from the first three generation quarks. 
Fig. \ref{fig:zprime} also shows the DM branching ratio is larger for $g_X=0.6$ than $g_X=0.3$.
This is because a larger coupling constant separates the DM mass from the $Z'$ boson resonance further (i.e. makes the DM mass smaller) in order to satisfy the relic density constraint.


\end{document}